# Numerical Analysis of Magnetic Domain Energy System in Nano-scale Fe

Shuji OBATA

*School of Science & Engineering, Tokyo Denki University, Hatoyama, Hiki, Saitama 350-0394, Japan*

Magnetic materials generally construct magnetic domains in external field $H$. These domain structures are changed with the field changes $\Delta H$ accompanying the Barkhausen effects. These phenomena are shown using Fe domain energy systems composed of classical magnetic dipole moment interactions. The magnetization curves are created with terraces and jumps, where the flux structure changes produce the Barkhausen noise. The terraces indicate the delays of the magnetization processes for the field $H$. These numerical simulations are performed in Fe nano-scale regular lattice systems of rods and belts, which directly show the evidences of these basic phenomena.

## 1. Introduction

Nano-scale magnets are variously proposed by a large number of processes using high technologies, which show various structures and properties.[1)-5)] The strong coercivity $H_c$ and the high remanent magnetizations $B_r$ are observed in recent experiments of nano-scale magnetizations. These results indicate new magnetic states and domain systems different from the general concepts for bulk states. As for Fe metal of soft magnetic materials, the nanowires of Fe produce the coercivity of $H_c$=450~600 Oe =4.0~4.8 ×10$^4$ A/m, which are created in carbon nanotubes.[1)] Non-saturate magnetization curves are also observed at room temperature in this system. Single domain Fe nanoparticle layers show the $H_c$=500 Oe : $\mu_0 H_c$=0.05 T,[2)] where the atomic magnetic moment is observed as $n_b\mu_B$: $n_b$=2.1 ($n_b$ and $\mu_B$ are the atomic spin number and the Bohr magneton respectively). In the report of Fe: Al$_2$O$_3$ nano-composite films, the value of magnetization per Fe atom is observed as a function of the Fe concentration from $n_b$=1.4 to 2.2 in a field of 2 T at 10 K, where the coercivity is shown as $H_c$=500 Oe.[3)] The low values of $n_b$=1.4 correspond to the slow saturation characteristics of nanoparticle magnetizations.

Analytical performances of a computer simulation method based on the classical magnetic dipole moment interactions are already confirmed in Refs. 6) and 7). The above experimental data are quantitatively analyzed using this method in this paper. For the Fe nano-scale systems, the almost same coercivities $H_c$=450~510 are obtained in our simulations. The energy minimum states calculated using whole moment interactions in the system directly show the domain structures and Barkhausen effects in Fe nano-scale BCC lattices. The Barkhausen effects are produced in the magnetization curves composed of terraces $\Delta H$ and jumps $\Delta M$ in the external field changes. These terraces and jumps are precisely obtained in the computer simulations, which clearly explain the experimental data. However, these simulations have not been correctly reported until now, because the calculations of the whole system interactions require the large scale computing resources in these days.[8)-11)] These calculations are impossible till 10 years ago in general circumstances, and finite temperature simulations using a Monte Carlo method require the largest scale computer system in present.

The simulation processes and the details of the minimum domain energy states are discussed in §2 for explaining the domain constructions. The magnetization energies are equated in §2.1 based on the classical magnetic dipole moment interactions. The moment energies of the parallel directions and the cross directions are calculated to find the energy minimum state with shifting the positions around various length nano-rod domains in §2.2. In §3, numerical simulations are performed using a nano-belt regular lattice of Fe. The domain structures and the energy systems are concretely represented in §3.1~ §3.3. The B effects are shown as the only domain structure transitions, which change flow out fluxes.

## 2. Magnetic Dipole Moment Interactions and Domain Energies
### 2.1 Magnetic dipole moment interactions

The B effects[12)-17)] and the domain structures[18)-22)] are investigated variously in various materials. Most of these results could be explained by simulations of the domain energy system composed of the atomic dipole moment interactions in the classical theory. The *B-H* characteristics are calculated with long range interactions between these magnetic dipole moments $\boldsymbol{\mu}_i$ and $\boldsymbol{\mu}_j$ localized at sites *i* and *j* respectively. The Bohr magneton

$$\boldsymbol{\mu}_B = \frac{eh}{4\pi m} = 9.274 \times 10^{-24} \text{ [A.m}^2\text{] [J/T]}$$

becomes the element of these magnetic moments, where constants are $\mu_0$=4π×10$^{-7}$ [N/A$^2$], $e$=1.602×10$^{-19}$ [C], $h$=6.626×10$^{-34}$ [J.s] and $m$ =9.109×10$^{-31}$ [kg]. The atomic magnetic moment [J.m/A] [Wb.m] is replaced by the dipole moment with small distance vector $\boldsymbol{\delta}$ [m] and magnetic charges $\Delta\varphi$ or $\Delta\psi$ [Wb] as

$$\boldsymbol{\mu}_B^0 = \mu_0 \boldsymbol{\mu}_B = \boldsymbol{\delta}\Delta\varphi, \quad \boldsymbol{\mu}_i = n_b \boldsymbol{\mu}_B^0 = \boldsymbol{\delta}_i \Delta\psi, \tag{1}$$

where $n_b$ is the effective number of spins in atoms. In a body center cubic lattice, dipole moment directions distribute variously in thermal fluctuations, where these regular type directions are drawn as A, B and C in Fig. 1. Arbitrary dipole moment interactions of these are shown in Fig. 2, where typical structures take parallel and cross moment directions in ground states.

Fig. 1.

Fig. 2.

Setting the distance vector $d_{ij}=e_{ij}d_{ij}$ between the dipole moments at $i$ and $j$ sites, the moment interaction energies are equated using the Taylor expansion of $(d \pm \delta)^{-1/2}$ as

$$W_{f,ij} = \frac{1}{4\pi\mu_0 d_{ij}^3}\{(\boldsymbol{\mu}_i \cdot \boldsymbol{\mu}_j) - 3(\boldsymbol{\mu}_i \cdot \boldsymbol{e}_{ij})(\boldsymbol{\mu}_j \cdot \boldsymbol{e}_{ij})\}. \qquad (2)$$

Regular lattice Fe takes the BCC structure with the lattice constant $a=2.86\times10^{-10}$ [m] at low temperature below 911 ℃ and have 2 atoms per a unit lattice. Now, the distance $d_{ij}$ is represented using coefficients $c_{ij}$ and the lattice constant $a$ as

$$d_{ij} = ac_{ij}. \qquad (3)$$

Setting $\boldsymbol{u}_i$ to the normal vector of the moment vector $\boldsymbol{\mu}_i$ at site $i$, the interaction energy $W_{f,ij}$ are represented between the moments $\boldsymbol{\mu}_i$ and $\boldsymbol{\mu}_j$ as like

$$W_{f,ij} = \mu_0 \frac{n_B^2 \mu_B^2}{4\pi d_{ij}^3}\{(\boldsymbol{u}_i \cdot \boldsymbol{u}_j) - 3(\boldsymbol{u}_i \cdot \boldsymbol{e}_{ij})(\boldsymbol{u}_j \cdot \boldsymbol{e}_{ij})\} = E_a f_{ij}, \qquad (4)$$

$$f_{ij} = \{(\boldsymbol{u}_i \cdot \boldsymbol{u}_j) - 3(\boldsymbol{u}_i \cdot \boldsymbol{e}_{ij})(\boldsymbol{u}_j \cdot \boldsymbol{e}_{ij})\}/c_{ij}^3, \qquad (5)$$

$$E_a = \mu_0 \frac{n_B^2 \mu_B^2}{4\pi a_{IS}^3} \to 7.166\times10^{-26} \quad J: \quad n_B = 2.22. \qquad (6)$$

The position factors $f_{ij}$ include 4 terms for two BCC lattice points. The structure energy $W_{f,j}$ [J] about the $j$-th dipole moment in a system becomes

$$f_j = \sum_i f_{ij}, \quad W_{f,j} = E_a f_j,$$

$$f = \sum_{i<j} f_{ij}, \quad W_f = E_a \sum_{i<j} f_{ij}, \qquad (7)$$

where the total counts take the summations $i<j$ evaluating the half count of $i, j$. The diagonal parts indicate the terms in unit lattices. The field-moment energies are equated in the linear superposition theory as

$$W_{H,j} = \boldsymbol{\mu}_j \cdot \boldsymbol{H}, \quad W_H = \sum_j \boldsymbol{\mu}_j \cdot \boldsymbol{H}. \qquad (8)$$

The atomic dipole moments $\boldsymbol{\mu}_i$ in a domain make the magnetization field $\boldsymbol{M}$ [T] for the moment $\boldsymbol{\mu}_j$ as like

$$\boldsymbol{M}_j = -\sum_i \frac{n_B \mu_B^0}{4\pi d_{ij}^3}\{\boldsymbol{u}_i - 3(\boldsymbol{u}_i \cdot \boldsymbol{e}_{ij})\boldsymbol{e}_{ij}\}. \qquad (9)$$

Thus, in external field $\boldsymbol{H}$ [H/m] by current $\boldsymbol{I}$ [A], the total domain energy $W$ is equated as like

$$W = W_H + W_f = \sum_j W_j$$

$$= -\sum_j \boldsymbol{\mu}_j \cdot (\boldsymbol{H} + \frac{1}{2} \boldsymbol{M}_j / \mu_0) = -\boldsymbol{M} \cdot \boldsymbol{H} + E_a f, \qquad (10)$$

where the magnetization field $\boldsymbol{M}$ is composed of the magnetic dipole moment arrays. Here, it becomes the important key point that the factor $f$ depends on the shape of the system in nano-scale. The magnetization $M_z$ in the field $H_z$ is estimated as the $z$ component of the total dipole moments:

$$M_z = (\sum_j \boldsymbol{\mu}_j)_z. \qquad (11)$$

The lowest energy state calculated in (10) constructs the domains composed of the same dipole moment clusters as in Fig. 3, where the domains make a looped flux structure in an aria. In Fe systems, the domains in a common flux loop aria produce the square loop structure as the energy minimum state. These looped domains close the fluxes inside the magnet body. The break downs of these or transitions to another looped structures produce the flow out fluxes as the B effect with the external field change $\Delta H$. These energy systems of the dipole moment interactions are analyzed in following sections.

Fig. 3.

*2.2 Energy factors between two lattice points*

As for Fe, the atomic spin states exist as the classical magnetic dipole moments in the quantum eigen-states. This fact has not been proved, nevertheless the calculated results show the good agreement with the experimental data.[1)-5)] Eqs. (1)-(11) represent the interaction energy system caused by the Bohr magneton, and the magnetization energies are equated using these classical dipole moment interactions. Although domain energies using the classical theory have been discussed from nearly 50 years ago, very few concrete works are reported under a few computational abilities.[8)-10)] In some magnetism, local spins interact through electrons in a conduction band, which do not behave as simple dipole moments. In such case, the magnetizations are investigated with different approaches such as the Stoner model as *s-d* coupling orbitals states.[24)]

Here, the stable energy states are estimated using eqs. (1)-(10) in a BCC lattice. The unit vector $\boldsymbol{u}_i$ and the body center position vector $\boldsymbol{s}_i$ are set as

$$\boldsymbol{u}_i = u_{ix}\mathbf{i} + u_{iy}\mathbf{j} + u_{iz}\mathbf{k}, \qquad (12)$$

$$\boldsymbol{s}_i = a(\mathbf{i}+\mathbf{j}+\mathbf{k})/2. \qquad (13)$$

According to eq. (2) and setting $s=1/2$, the position vectors $\boldsymbol{r}_{\alpha\beta}$ are defined for 4 atoms $A_i$, $a_i$ and $B_j$, $b_j$ as $\alpha \in A, a$ and $\beta \in B, b$ at $i$ and $j$ lattice points;

$$\boldsymbol{r}_{BA}(k,l,m) = \boldsymbol{r}_{ba} = a(k\mathbf{i}+l\mathbf{j}+m\mathbf{k}), \qquad (14)$$

$$\boldsymbol{r}_{Ba}(k,l,m) = a\{(k+s)\mathbf{i}+(l+s)\mathbf{j}+(m+s)\mathbf{k}\}, \qquad (15)$$

$$\boldsymbol{r}_{bA}(k,l,m) = a\{(k-s)\mathbf{i}+(l-s)\mathbf{j}+(m-s)\mathbf{k}\}. \qquad (16)$$

The atomic dipole moment takes energy states for each direction around some directed domain structure. These energy calculations are very important to know the mechanism of the domain structure constructions. In this paper, two typical energy states of parallel and cross moment directions are calculated to find the lowest energy states briefly. In Fig. 2, the restricted relations between parallel $\boldsymbol{u}_p$ and the cross $\boldsymbol{u}_c$ dipole moments are equated using the angle $\theta_j = \theta_i \text{ or } (\theta_i - \pi/2) \rightarrow \theta_i = \theta_{\alpha\beta}$ as

$$\cos\theta_{\alpha\beta}(k,l,m) = (\boldsymbol{u}_p \cdot \boldsymbol{e}_{\alpha\beta}\cdot), \qquad (17)$$

$$\sin\theta_{\alpha\beta}(k,l,m) = (\boldsymbol{u}_c \cdot \boldsymbol{e}_{\alpha\beta}\cdot). \qquad (18)$$

Energy structure factors $f_j$ in (7) of a magnetic dipole moment are calculated for nano-rod type domains which center is set to the coordinate origin O. In these BCC lattice calculations, the lowest energy of dipole-domain interactions appears in parallel or cross moment direction according to the domain positions. These energy mechanisms are investigated using the structure factors $f_j$ at various positions in next section, where only checks in $X$-positions are introduced in Ref. 6). Practical energy values about one moment are determined by several environment domains. These actual steady states are calculated in §3 as domain structuring simulations using the 26 directions.

*2.3 Parallel and cross dipole moment energy factors*

Generally, the thermal fluctuations destroy the dipole moment fields. Thus, the systems are assumed to be in low temperature in this section. The basic energy systems between the dipole moments and the magnetic domains are calculated for clearing the domain structure constructions. The energy factors $f_{ij}$ in (7) are estimated between the moments at a lattice point $j(n_k, n_l, n_m)$ and other points $i(k, l, m)$ in rod type domains in Fig. 4 and Fig. 5.

Fig. 4.

Fig. 5.

The Ferro-stats are calculated on a $X$-coordinated direction as in Fig. 1-A as a ground state. The factor $f^F$ is set to merely $f$. Now, we consider magnetizations in nano-rod type domains directed to $Z$ axis with sizes

$$N=(2n_x+1)(2n_y+1)(2n_z+1), \qquad (19)$$

where $n_x$, $n_y$ and $n_z$ are half size of the domain rods. For estimating the energy factors, the full lattice points of $i$ in these domains are counted by $k$, $l$ and $m$ for each sampling moment $\boldsymbol{\mu}_j$ at the position $n_k$, $n_l$ and $n_m$:

$$-n_x \leq k-n_k \leq n_x, \quad -n_y \leq l-n_l \leq n_y, \quad -n_z \leq m-n_m \leq n_z. \qquad (20)$$

In Fig. 6, the parallel moment energy factors $f_{0//}$ at the domain center are represented as the typical structure. The parallel moment energy factors $f_{j//}$ traced on $X$ axis and $Z$ axis in Fig. 4 are represented in Fig. 7 Trace{I} and Trace {II} respectively. The cross moment energy factors $f_{j\perp}$ traced on $X$ axis and $Y$ axis on the domain edge surface in Fig. 5 are represented in Fig. 8 Trace {III} and Trace {IV} respectively.

Fig. 6.

Fig. 7.

The important results of the parallel moment energy factors $f_{j//}$ in rod type domains are clarified as follows. ①The moments in short domains are quite unstable with large + values as like at $n_z=5$ for $n_x=n_y>6$ in Fig. 6. ②The middle part moments in long domains are stable as like at $n_z=20$ for $n_x=n_y<11$ in Fig. 6. ③The sign of near the side edges becomes − in long domains, and this sign changes drastically to + at the edge points ($n_x \to n_x+1$) as shown in Fig. 7 (c) and (d). This means that the adjacent domains at sides become stable with anti-parallel moments. ④The + sign appears near the cross section edges, and the sign changes drastically to − at the edge points ($n_z \to n_z+1$) as in Fig. 7 (g) and (h). These results mean that the docking of small domains produces stable state and increases continued long domains.

Fig. 8.

The important results of the cross moment energy factors $f_{j\perp}$ on the edge surface in Fig. 8 are clarified as follows. ⑤In Fig. 8 Trace {III} (*l*); $n_k=-(n_x+1)$, the factors take the energy minimum values. This means that near the rod domain edges, adjacent domains become stable with the cross moment and should be going to make the looped flux structures. ⑥All of the cross moment factors have the − sign to be stable, which make the looped structure domains.

In the previous work in Ref. 6), it is clarified that the energy factors of $f_{ij}$ between magnetic dipole moments $f_b$ and $f_c$ with the directions in Fig. 1 B and C become all smaller than $f_a$ of X, Y and Z axis directions in Fig. 1 A. The energy factors of $f_b$ and $f_c$ with parallel and cross moment directions produce the weaken values in oblique rod systems as like

$$e_P=(\mathbf{j}+\mathbf{k})/\sqrt{2},\ e_C=\mathbf{i} \qquad \to f_b \approx 0.91 f_a \qquad (21)$$

$$e_P=(\mathbf{i}+\mathbf{j}+\mathbf{k})/\sqrt{3},\ e_C=(-\mathbf{i}+\mathbf{j})/\sqrt{2} \to f_c \approx 0.82 f_a. \qquad (22)$$

These decay factors in Fe are not so large compared with the energy factors of the domain constructions. General domain structures depend on such easy magnetization axes, and show quite different configurations according to these decay factors. Relations of the domain structures and the easy axes are also complex in polycrystalline including grain boundaries.[25]

## 3. Domain structure and the Barkhausen Effect
### 3.1 Magnetic dipole moment array

We can directly simulate the Fe magnetizations in nano-scale regular lattice systems using the energy equations in §2.1, where the magnetization characteristics nicely confirm with experimental data in nano-scale materials.[1)-5)] Based on the energy constructions of the dipole moments, the domain structures are analyzed precisely and visually in this section. The simulations are performed in a nano-belt system under the 26 moment directions in Fig. 9.

Fig. 9.

Setting the values $e_1=1$, $e_2=1/\sqrt{2}$ and $e_3=1/\sqrt{3}$, the 26 directions are represented using the marks as '>': ($e_1$,0,0), '+': (0,$e_1$,0), '1': (0,0,$e_1$), '-': (-$e_1$,0,0), '•': (0,-$e_1$,0), 'W': (0,0,-$e_1$), 'g': ($e_2$,$e_2$,0), 'h': (0,$e_2$,$e_2$), '^': ($e_2$,0,$e_2$), 'j': (-$e_2$,$e_2$,0), 'k': (0,-$e_2$,$e_2$), '`': (-$e_2$,0,$e_2$), 'm': ($e_2$,-$e_2$,0), 'n': (0,$e_2$,-$e_2$), ']': ($e_2$,0,-$e_2$), 'p': (-$e_2$,-$e_2$,0), 'q': (0,-$e_2$,-$e_2$), '(': (-$e_2$,0,-$e_2$), 's': ($e_3$,$e_3$,$e_3$), 't': (-$e_3$,$e_3$,$e_3$), 'u': ($e_3$,-$e_3$,$e_3$), 'v': ($e_3$,$e_3$,-$e_3$), 'w': (-$e_3$,-$e_3$,$e_3$), 'x': (-$e_3$,$e_3$,-$e_3$), 'y': ($e_3$,-$e_3$,-$e_3$), 'z': (-$e_3$,-$e_3$,-$e_3$). The main marks are shown in Fig. 9 (b). The energy minimum states are calculated by means of these directed dipole moment interactions in a $N_x N_y N_z$=20×4×40 lattice system with $n_b$=2.22. The obtained moment directions are drawn using these marks to show the domain structures. The first step random array and the annealed state in the initial trace curve at $H$=5×10$^4$ A/m are shown in Fig. 10.

Fig. 10.

The simulations are performed using following 4 processes. **A**: Random state start. At first, the above 26 directions are randomly set for all sites. **B**: External field trace. In this paper, a linear field trace is adapted using the values of integer $k$, $H_m$=1×10$^5$ A/m and $K$=100 as

$$H_k = H_m \frac{k}{K}, \quad 0 \leq k \leq K, \quad K \geq k \geq -K, \quad -K \leq k \leq K. \qquad (23)$$

In the precise calculations, the division is set to $K$=1250 and the concerned data are picked up in the third trace. **C**: Cooling. Cooling is performed as almost 0 K using a direct energy cut off method instead of a usual Monte Carlo method. This process is executed in *X-Z* plane traces of sites *j* from the front surface to the back surface with taking the energy minimum state of $W_j$ in (10) under full summations of the other sites *i*. The energies are checked in 26 directions and the minimum energy state is selected at every site. This process is contentiously performed with tracing the *X*, *Z* and *Y* coordinate step by step. The summations of $f_{ij}$ in (7) are counted over full system moments. These processes are executed through *n* time iterations till ($W_{j,n}$-$W_{j,n+1}$)/$W_{j,n}$ < 10$^{-4}$, where the converging process is performed with almost *n*~5 iterations under *n*<40. **D**: Representation. The dipole moment directions are printed using the above 26 marks as the energy minimum structures, where the looped domain structures are observed clearly. The dipole moment energies are represented using a density plot method with the intensity from black of the minimum value to white of the maximum value.

In the cooling processes, there are important comments as follows. **I**: The Monte Carlo method based on the replacement with the Boltzmann factor probability is very slow for this cooling. **II**: The temperature of the domain structures should be set based on averaged value of gained energies from the grand state. This determination could not be performed with usual methods, where the difficulty lurks in aberrance transitions of domains deviating from the thermal equilibrium states. **III**: There is an important question that these domain transitions how to include the quantum phenomena under absolutely quantized fluxes.

*3.2 Magnetization curve and Barkhausen effect*

The aria composed of a same dipole moment direction is called the domain, and such magnetic domains construct looped flux structures through long range interactions. The magnetization curves are obtained in the nano-belt system under linear external field trace of $H_m=1.0\times10^5$ A/m as in Fig. 11. The maximum magnetization $B_m$ par atom is normalized to $n_b$. The magnetization curves confirm with the experimental data indicated in §1 as the coercivity $\mu_0 H_c=0.0456$ T, the remanent magnetization par atom $B_r=2.0$.

Fig. 11.

The terraces $\Delta H$ and the jumps $\Delta M$ are observed in precise calculations of magnetization curves. The sample at an aria A in Fig. 11 is drawn to show these B effects as in Fig. 12. This precise curve is different from the coarse one because of a short calculation using a different field trace. The terraces indicate that the domains do not break down at external field $H$ till some amount of added field $\Delta H$, where the jumps of $\Delta M$ make the B noises. These are observed as flux changes of $\Delta\varphi$ flowed out from a magnet body and are transformed to voltage $m\Delta\varphi/\Delta t$ in a $m$ turn coil.

Fig. 12

*3.4 System energy, domain patterns and energy distributions*

We can simulate the Fe regular-lattice magnetization in the nano-scale systems using Eq. (10) under nice confirmations with the many experimental data[1)-5)] introduced in §1. The simulations are performed in the nano-belt system shown in Fig. 9, where the magnetization curves include big transitions as in Fig. 11. The B effects are certified with the precise calculations of these curves including the terraces $\Delta H$ and the jumps $\Delta M$ shown in Fig. 12. These jumps accompany the domain structure changes. In this paper, new treatments are introduced for representing such domain changes using the dipole moment distributions compared with the energy distributions.

The domain energy $W$ is composed of two components as the field term $W_H=-\boldsymbol{M}\cdot\boldsymbol{H}$ and the structure term $W_f=E_a f$ in (10). These energy structures in the $20\times4\times40$ lattice point system are drawn in Fig. 13 about (a) the field term, (b) the structure term and (c) the total energy. These energies have the following characters in the field tracing. ① In decreasing field, these three energies commonly change with linear lines. ② In increasing field, these energies change variously accompanying jumps. ③ The field terms are weaker (larger) than the structure terms about $3.5\times 10^{-20}$ J. ④ In the field terms, the increasing trace are larger than the decreasing trace. ⑤ In the structure terms, the increasing trace is smaller than the decreasing trace. ⑥ The field terms in increasing trace sometimes take plus energy values. ⑦ In the total energy, two increasing traces are cancelled each other without jumping arias. ⑧ In low fields, the total energy take large values, where the moments store miss much energy stresses.

Fig. 13.

The domain structures and the moment energy distributions are drawn in Fig. 14 (a) (b) and (c), which are corresponded to figures at $J_a$, $J_b$ and $H_c$ points in Fig. 12 and ○ marks in Fig. 13 (c) respectively. The upside figures show the domain structures using the marks in Fig. 9 (b).

Fig.14

The downside figures show the energy distributions of $W_j$ in (10), where the darkness from the black to the white show the values from $-7.5\times 10^{-24}$ to $-2.5\times 10^{-24}$ J. The domain structure transitions are directly observed in the movements of the white arias representing the domain walls. The figures (a) and (b) represent the drastic domain transitions as an avalanche in the field change $\Delta H$=159 A/m (2 Oe).

As for the experiments, these local domain structure changes are realistically observed in permalloy thin-film microstructures,[23] which correspond to drastic jumps in the magnetization curves producing the giant B noise. These phenomena are observed in many experimental data in various nano-scale magnets.

## 4. Summary

The magnetization mechanisms in Fe are directly represented by the numerical simulations using the atomic magnetic dipole moment interactions under the classical theory. The energy system in the magnetization processes is clearly shown using the magnetization curves, the domain patterns, the energy curves of the dipole moments and these distribution patterns. The Barkhausen effects in the regular lattices are clarified with the precise calculations of the magnetization curves and with the domain break down structures.

**Acknowledgement**

This work is accomplished by using the supercomputer system in Institute of Solid State Physics, University of Tokyo. The author greatly thanks for ISSP computer center.

1) B.C. Satishkumar, A. Govindaraj, P.V. Vanitha, Arup K. Raychaudhuri, and C.N.R. Rao: Chem. Phys. Let. **362** (2002) 301.
2) K.D. Sorge, J.R. Thompson, T.C. Schulthess, F.A. Modine, T.E. Haynes, S. Honda, A. Meldrum, J.D. Budai, C.W. White, and L.A. Boatner : IEEE Trans. Magn., **37** (2001) 2197.
3) N.M. Dempsey, L. Ranno, D. Givord, J. Gonzalo, R. Sema, G.T. Fei, A.K. Petford-Long, R.C. Doole, and D.E. Hole: J. Appl. Phys, 90 (2001) 6268-6274.
4) M. Yoon, Y.M. Kim, Y. Kim, V. Volkov, H.J. Song, Y.J. Park, S.L. Vasilyak, and I.-W. Park: J. Magn. Magn. Mater. **265** (2001) 357-362.
5) Wen-Chin Lin, C.B. Wu, P.J. Hsu, H.Y. Yen, Zheng Gai, Lan Gao, Jian Shen, and Minn-Tsong Lin: J. Appl. Phys., **108** (2010) 034312.
6) S. Obata: IEEJ Trans. FM, **133** No 9 (2013) 489.
7) T. Koyama : Sci. Technol. Adv. Mater. **9** (2008) 013006_1-9.
8) S. Chikazumi: *Physics of Ferromagnetism* (2nd edn. Oxford: Oxford University Press, 1997) pp. 1-10.
9) S.L.A. de Queiroz and M. Bahiana : Phys. Rev. E. **64**, (2001) 066127-1-6.
10) H.T. Savage, D-X. Chent, C. Go'mez-Polo, M. Va'zquez, and M. Wun-Fogle : J. Phys. D; Appl. Phys. **27** (1994) 681-684.
11) H. Kronmüller, H.-R. Hilzinger, P. Monachesi, and A. Seeger : Appl. Phys. **18** (1979) 183-193.
12) S. Zapperi, P. Cizeau, G. Durin, and E. Stanley : Phys. Rev. B, **58** (1998) 6353-6366.
13) F. Colaiori, and A. Moro : Advances in Physics, **57** (2008) 287-359.
14) S. Yang and J.L. Erskine : Phys. Rev. B, **72** (2005) 064433-1-13.
15) K. Kova'cs, and Z. Ne'da : J. Opt. Elect. Adv. Mater, **8** (2006) 1093-7.
16) D.A. Christian, K.S. Novoselov, and A.K. Geim : Phys. Rev. B, **74** (2006) 06443_1-6.
17) G. Durin, and S. Zapperi : J. Stat. Mech., **2006** (2006) P01002_1-11.
18) D.C. Jiles : Czech. J. Phys., **50** (2000) 893-988.
19) O. Gutfleisch, K.-H. Müller, K. Khlopkov, M. Wolf, A. Yan, R. Schäfer, T. Gemming, and L. Schultz : Acta Materialia, **54** (2006) 997-1008.
20) T. Eimüller, P. Fischer, G. Schütz, P. Guttmann, G. Schmahl, K. Pruegl, and G. Bayreuther : J. Alloys and Compounds, **286** (1999) 20-25.
21) P. Fischer, M.-Y. Im, T. Eimüller, G. Schütz, and S.-C. Shin : J. Magn. Magn. Mater, **286** (2005) 311-314.
22) N.I. Vlasova, G.S. Kandaurova, and N.N. Shchegoleva : J. Magn. Magn. Mater, **222** (2000) 138-158.
23) S. Yang and J.L. Erskine : Phys. Rev. B, **72** (2005) 064433-1-13.
24) S. Wakoh, and J. Yamashita : J. Phys. Soc. Jpn., **21** (1966) 1712.
25) K. Sato, Y. Murakami, D. Shindo, S. Hirosawa, and A. Yasuhara: Materials Transactions, **51** (2010) 333.

**Figures**

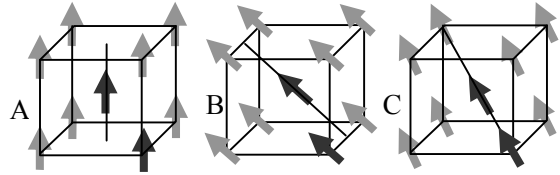

Fig. 1 Ferro magnetic moments in a BCC lattice. The dipole moment directions are mainly divided to 3 types of A, B and C. The domain energies have the largest value in the type A.

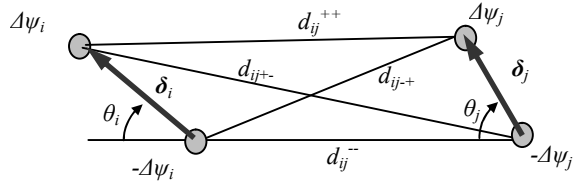

Fig. 2. Two magnetic moment interactions in arbitrary directions. The dipole moments are equated as $\boldsymbol{\mu}_i = n_b \boldsymbol{\mu}_B^0 = \boldsymbol{\delta}_i \Delta \psi$.

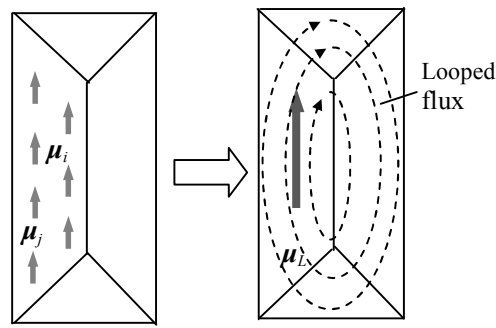

Fig. 3. Atomic dipole moments in an aria. A stable energy state aria makes a looped flux structure. These flux structures are the bases of the domain constructions.

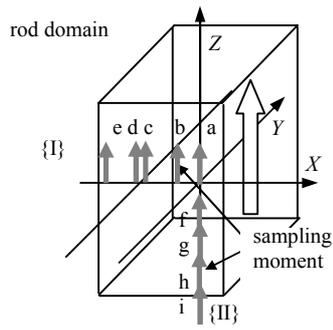

Fig. 4.　Dipole moment traces on the central axes. {I}: $X$ axis, {II}: $Z$ axis.

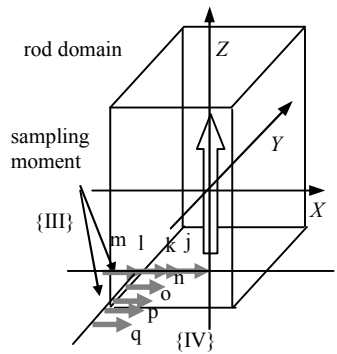

Fig. 5.　Dipole moment traces on axes. {III}: $X$ direction at $l=0$ and $m=-n_z-1$, {IV}: $Y$ direction at $k=-n_x-1$ and $m=-n_z-1$.

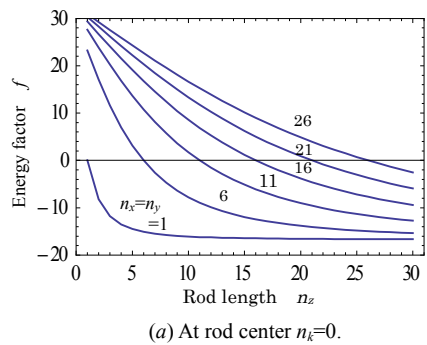

(*a*) At rod center $n_k=0$.

Fig. 6.　A moment energy factor of a *j* lattice point at the domain center. Main phenomena are included in this figure. It is directly shown that long rod domains are stable.

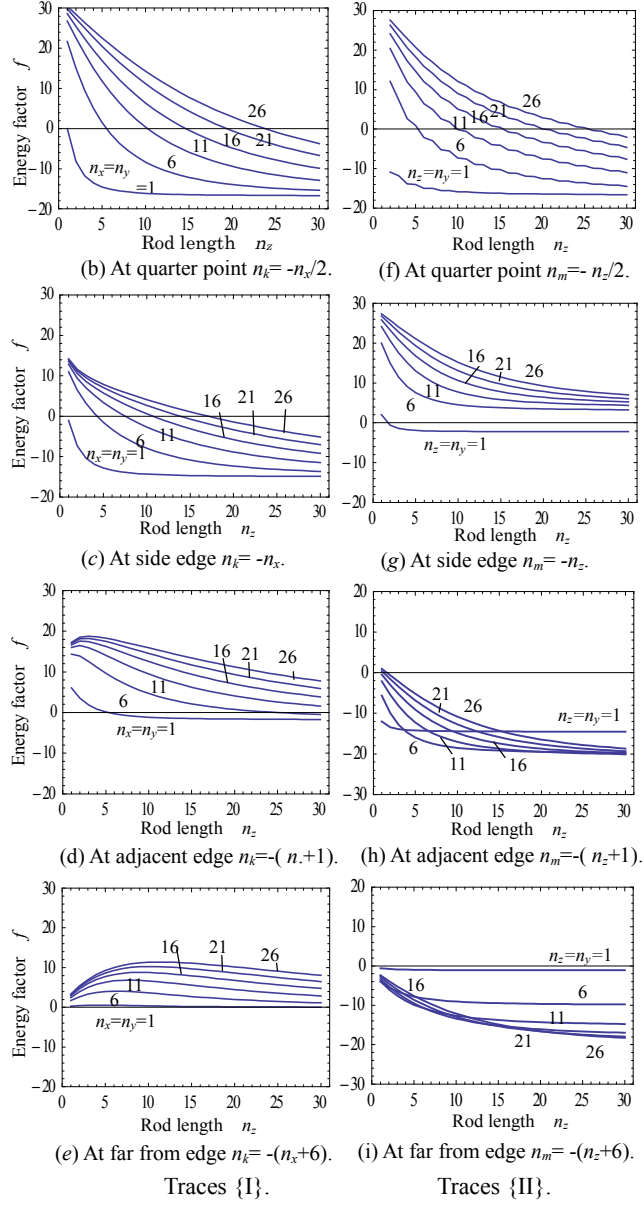

Fig. 7. Energy factors of parallel dipole moments for domains tracing on $X$ {I} and $Z$ {II} axes in Fig. 4.

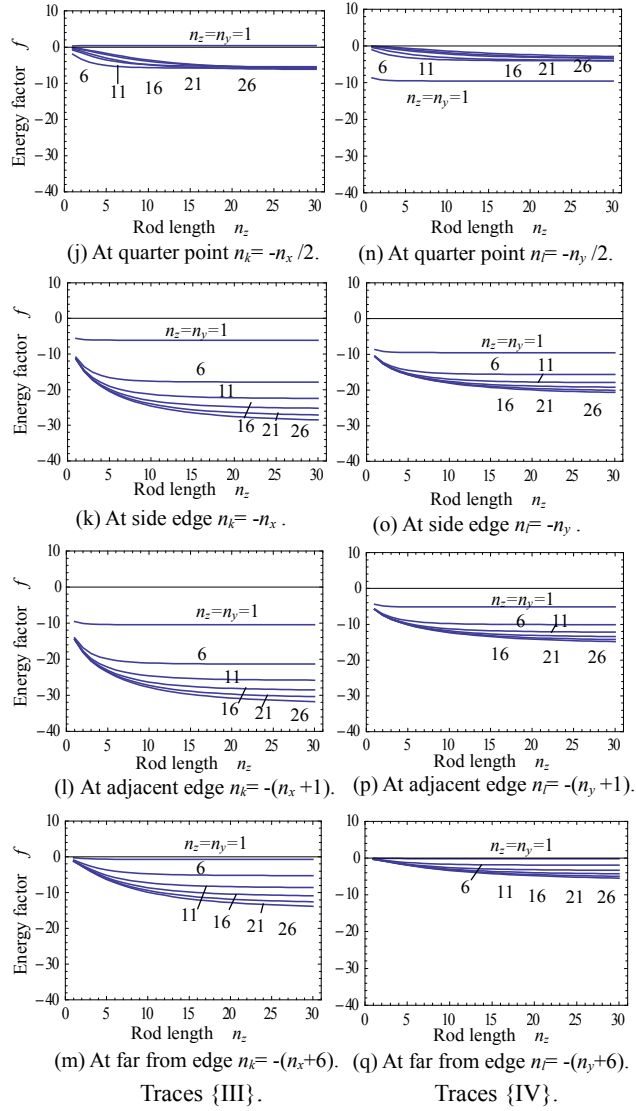

Fig. 8. Energy factors of cross direction dipole moments for rod domains tracing on $X$ {III} ($l=0, m=-n_z-1$) and $Y$ {IV} ($k=-n_x-1, m=-n_z-1$) axes in Fig. 5.

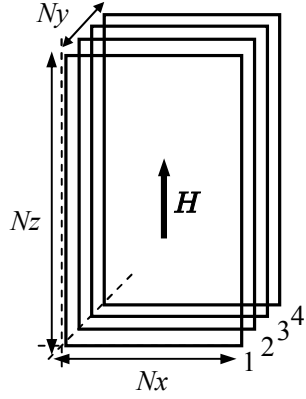

(a) System sizes and cut sheet number.

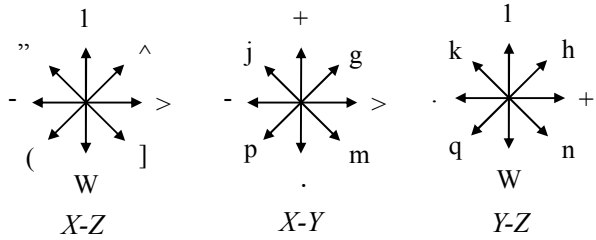

(b) Directions and marks in *X-Z*, *X-Y* and *Y-Z* planes.

Fig. 9. Nano-scale system size and moment direction marks. (a) The system scale of $N_x$, $N_y$ and $N_z$ lattice points and cut sheets in a orthorhombic system. (b) Main direction marks in *X-Z*, *X-Y* and *Y-Z* planes.

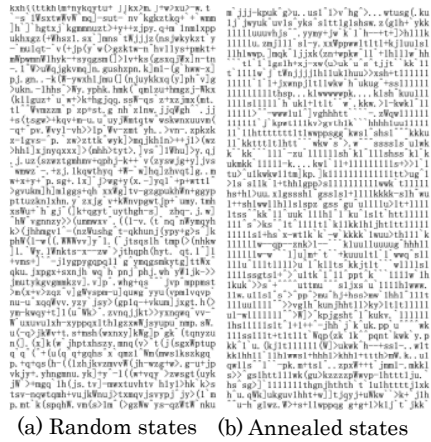

(a) Random states  (b) Annealed states

Fig. 10. Nano-belt dipole moment arrays at $y=2$ in a $N_xN_yN_z=20\times4\times40$ lattice point system with $n_b=2.22$. (a) The first random arrays. (b) The annealed moment arrays in the initial trace curve at $H=5\times10^4$ A/m and T<10 K.

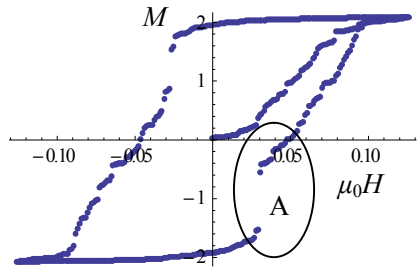

Fig. 11. Magnetization curve ($B_m=n_b$) in a $NxNyNz=20\times4\times40$ lattice point system. The aria A is precisely calculated and drawn in Fig. 12 to show the B effects.

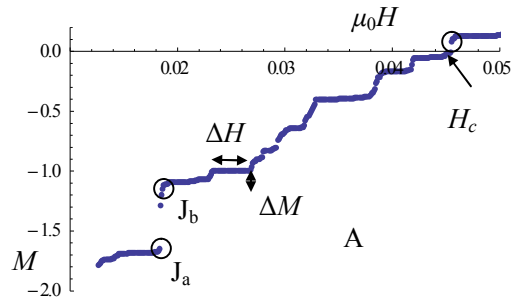

Fig. 12. The terraces $\Delta H$ and the jumps $\Delta M$ of the B effects in the precise magnetization curve at the aria A in Fig. 11. The coercivity becomes $\mu_0 H_c =0.0456$ T.

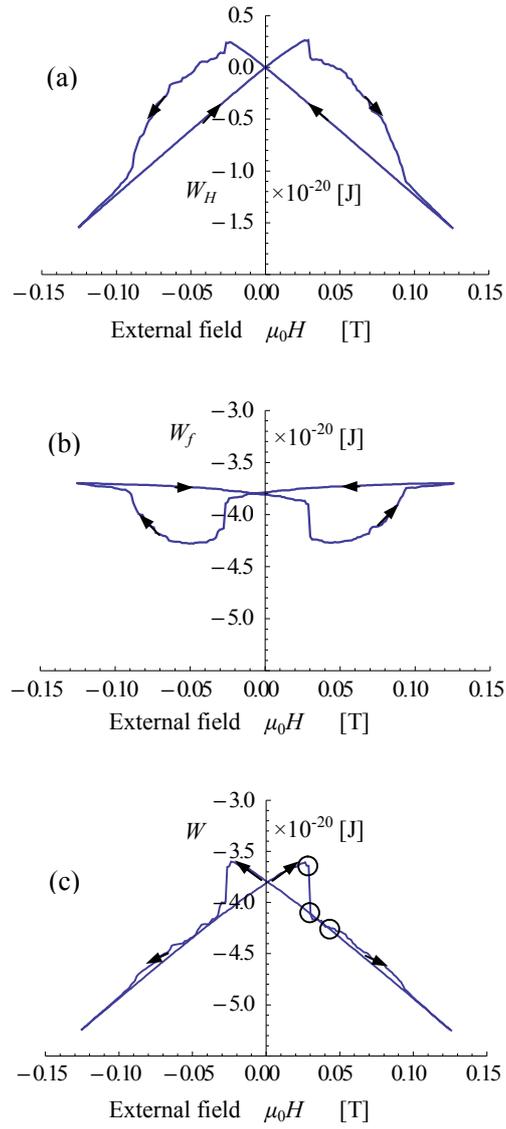

Fig. 13. System magnetization energies. (a) The variations of the moment-field energy: $W_H$. (b) The moment-structure energy: $W_f$. (c) The total energy: $W$. The energy curves are corresponded to the magnetization curves in Fig. 11. The ○ marks in (c) indicate the $J_a$, $J_b$ and $H_c$ points in Fig. 12 respectively.

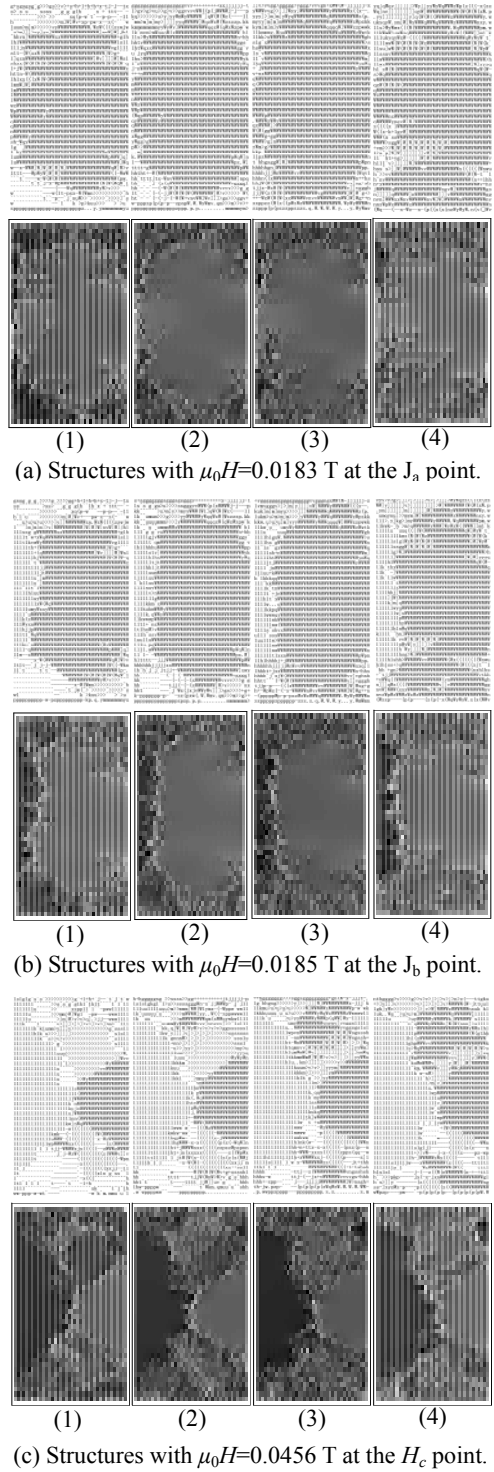

(a) Structures with $\mu_0 H=0.0183$ T at the $J_a$ point.

(b) Structures with $\mu_0 H=0.0185$ T at the $J_b$ point.

(c) Structures with $\mu_0 H=0.0456$ T at the $H_c$ point.

Fig.14.   Domain structures (upside) and energy distributions of $W_j$ (downside). The figures of (a), (b) and (c) are corresponded to the $J_a$, $J_b$ and $H_c$ points in Fig. 12 and the ○ marks in Fig. 13 (c) respectively. The numbers of (1)~(4) indicate the sheet positions in $y$-axis.